\documentclass{article}
\usepackage[dblblindworkshop, final]{neurips_2026}
\workshoptitle{Linguistic Principles for Foundation Models (LP4FM)}
\usepackage[utf8]{inputenc}
\usepackage[T1]{fontenc}
\usepackage{hyperref}
\usepackage{url}
\usepackage{booktabs}
\usepackage{amsfonts}
\usepackage{amsmath}
\usepackage{amssymb}
\usepackage{amsthm}
\usepackage{nicefrac}
\usepackage{microtype}
\usepackage{xcolor}
\usepackage{array}
\usepackage{calc}
\usepackage{colortbl}
\usepackage{enumitem}
\usepackage{graphicx}
\usepackage{mathtools}
\usepackage{multirow}
\usepackage{subcaption}
\usepackage{cleveref}
\usepackage{wrapfig}

\title{Beyond Structural Unification: A Perspective and Roadmap on Unified Multimodal Models}

\author{%
  \textbf{Yinyi Luo\textsuperscript{1,*} \quad
  Hayes Bai\textsuperscript{2,*} \quad
  Yiqiao Jin\textsuperscript{3,*} \quad
  Wenyue Hua\textsuperscript{4}} \\
  \textbf{Srijan Kumar\textsuperscript{3} \quad
  Marios Savvides\textsuperscript{1} \quad
  Qingsong Wen\textsuperscript{5} \quad
  Jindong Wang\textsuperscript{2}} \\
  \textsuperscript{1}Carnegie Mellon University \quad
  \textsuperscript{2}William \& Mary \quad
  \textsuperscript{3}Georgia Institute of Technology \\
  \textsuperscript{4}Microsoft Research \quad
  \textsuperscript{5}Squirrel Ai Learning \\
  \textsuperscript{*}Equal contribution.
}

\begin{document}
\maketitle

\begin{abstract}
Unified Multimodal Models (UMMs) emerge as a promising way to combine understanding and generation within a single architecture.
We argue that current approaches mainly achieve structural unification, which can create a \emph{Unification Paradox}: shared interfaces, token spaces, and parameters do \textit{not} guarantee an overall capability gain and may degrade individual capabilities.
In this position paper, we identify five critical layers in which unification must be rethought: (1) \textit{Asymmetric Input Unification}, where models must estimate the utility and reliability of each modality rather than treating all modalities equally; (2) \textit{Factorized Representation Unification}, where semantic abstraction and fine-grained reconstruction should use separate but coordinated capacity; (3) \textit{Stateful Reasoning Unification}, where models should maintain and revise a shared multimodal belief state across understanding and generation; (4) \textit{Synergistic Optimization Unification}, where conflicting understanding and generation objectives require coordinated optimization; and (5) \textit{Multi-Centric Systemic Evaluation}, where benchmarks must evaluate systemic competence, capability preservation, and cross-task robustness rather than isolated outcomes.
We argue that progress requires moving beyond structural unification toward functional unification across the full system.
\end{abstract}
\vspace{-10pt}

\section{Introduction}
\vspace{-10pt}

\begin{wrapfigure}{r}{0.5\textwidth}
    \centering
    \vspace{-20pt}
    \includegraphics[width=0.45\textwidth]{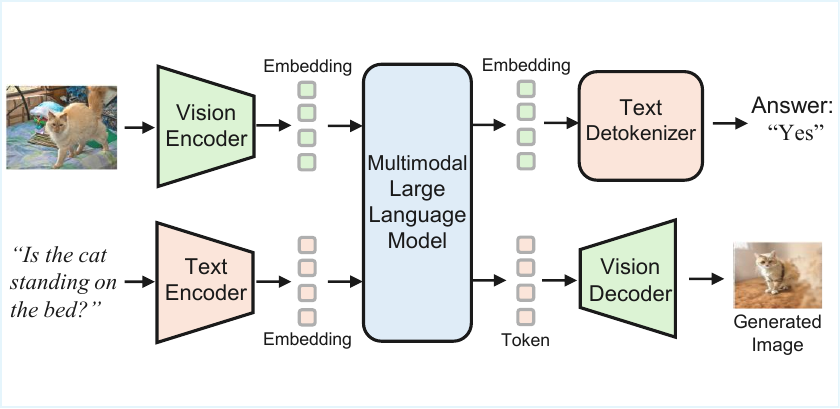}
    \vspace{-.2in}
    \caption{A general architecture of UMM.}
    \label{fig-umm}
    \vspace{-.5in}
\end{wrapfigure}

Unified Multimodal Models (UMMs) \citep{deng2025emerging, ma2025janusflow, xie2024show, chen2024janus} integrate perception, reasoning, and generation within a single network (\Cref{fig-umm}). By collapsing modal processors into a shared backbone, they promise transfer and coordination beyond modular pipelines. This has encouraged a \textit{structural unification paradigm}: tokenizing pixels, sharing interfaces, and co-locating parameters are often treated as proxies for systemic synergy.

Existing UMMs pursue this paradigm through diffusion-based co-location~\citep{swerdlow2025unified, yang2025mmada}, autoregressive interface homogenization~\citep{team2024chameleon, wang2024emu3, sun2024generative, ge2024seed, chen2025janus}, and fused architectures that stitch autoregressive decoding with diffusion or flow-based visual synthesis~\citep{xie2024show, ma2025janusflow, deng2025emerging}. These designs achieve remarkable engineering elegance, but they still prioritize structural and code-level interface unification over functional and behavioral coherence.

\begin{figure}[t!]
    \centering
    \includegraphics[width=.8\linewidth]{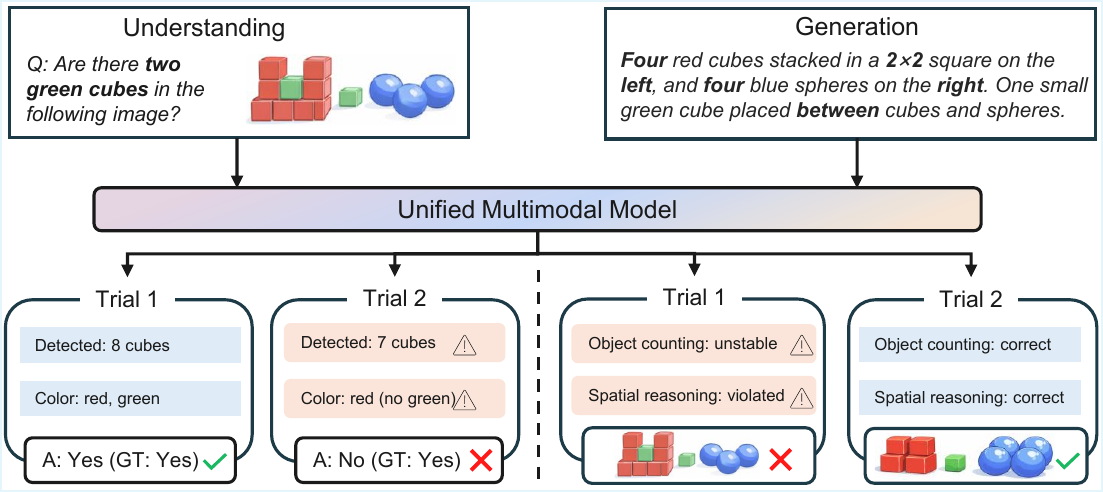}
    \vspace{-.1in}
    \caption{An illustration of Unification Paradox.}
    \label{fig-motiv}
    \vspace{-.2in}
\end{figure}

However, structural consolidation does not by itself establish functional coordination. We call this gap the ``Unification Paradox'': a more structurally unified model can still exhibit capability interference, inconsistent reasoning, or degradation in an individual function. As illustrated in \Cref{fig-motiv}, a model may solve a visual puzzle in one trial but fail in an isomorphic setting, or generate a high-fidelity image that violates its own spatial constraints \citep{zhou2025multi, xu2026towards, team2024chameleon}. TorchUMM~\citep{luo2026torchumm} provides concrete examples under a standardized evaluation interface. Supervised fine-tuning of Bagel raises its reported MMMU score from $0.519$ to $0.526$ but lowers MM-Vet from $65.9$ to $61.2$; on TokenFlow, the same class of intervention lowers DPG-Bench from $71.29$ to $22.16$. Its controlled case studies also find that a less structurally unified system can outperform more tightly shared systems on coupled understanding--generation tasks. These results do not isolate structural sharing as the sole cause, because the compared models differ in data, initialization, and generative formulation. They do demonstrate the observable failure mode: a local gain or a more unified design can coexist with material movement elsewhere in the capability profile. We therefore treat them as motivating quantitative evidence rather than a universal causal law.

We therefore define the goal of multimodal integration as a systemic gain subject to capability-preservation constraints, rather than interface sharing alone. The Unification Paradox can arise through coordination gaps across five layers of the UMM stack:
(1) Input Layer (Asymmetric Input Unification): Multimodal inputs possess intrinsic asymmetries in evidence reliability and information density. Current UMMs force them through identical sequence lengths or fixed structural pipelines, which paradoxically allows noisy or low-quality modalities to corrupt clean internal states, inducing overconfident hallucinated behaviors \citep{zollo2025confidence, zitkovich2023rt}.
(2) Representation Layer (The Capacity Allocation Paradox): Understanding favors high-level semantic abstraction, whereas generation requires fine-grained reconstructive detail. A monolithic latent space can therefore create capacity encroachment and negative transfer between the two objectives.
(3) Reasoning Layer (Stateful Reasoning Disregard): Current cross-modal reasoning is predominantly text-centric, treating visual tokens as passive static conditioning signals. The model lacks a unified, dynamic internal belief state to maintain multi-turn or cross-modal logical coherence, leading to severe attribute mixing and compositional failure during complex long-horizon decisions \citep{tang2025unilip}.
(4) Post-Training Layer (Objective Cannibalization): The joint training of discriminative and generative paradigms relies on aggressive multi-task text-stitching. Without explicit gradient harmonization and Pareto-optimal reward shaping, the training objectives fiercely collide, tearing apart the model's internal consistency.
(5) Evaluation Layer (The Single-Benchmark Illusion): Community assessments rely heavily on decoupled, outcome-oriented leaderboards (e.g., measuring understanding and generation through isolated metrics like VQA accuracy and FID). This single-benchmark optimization illusion can mask collateral capability regressions inside structurally unified models, making local gains appear stronger than the unified system they are meant to evaluate.

\begin{figure}[t!]
    \vspace{-20pt}
    \centering
    \includegraphics[width=\linewidth]{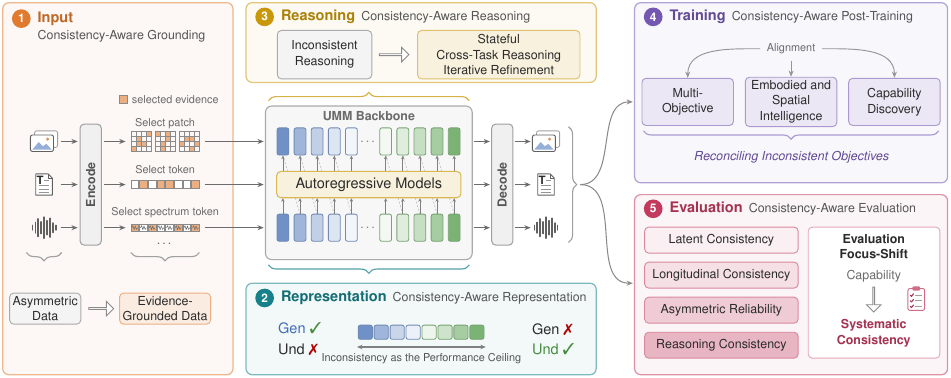}
    \caption{We argue that systemic competence should supersede structural unification as the priority for next-generation UMMs.}
    \label{fig:overall}
    \vspace{-0.1in}
\end{figure}

Our position is that systemic competence and functional unification must supersede mere structural homogenization as the guiding principle for next-generation UMMs (\Cref{fig:overall}). True unification cannot be inferred from code-level and interface-level homogenization alone; it truly requires coordinated input calibration, factorized representation learning, stateful cross-modal reasoning, synchronized optimization, and rigorous systemic evaluation. The five layers are deliberately scoped to the interfaces at which evidence enters, is represented, is updated, is optimized, and is audited. They are not intended as an exhaustive taxonomy of multimodal modeling. Through this lens, we chart a falsifiable path toward \textit{coherent unified multimodal intelligence}, where capabilities are not merely structurally co-located, but are jointly improved or demonstrably preserved.

\section{Position Statement}
\label{sec-position}
\vspace{-10pt}

We distinguish two claims that are often conflated. A model is \emph{structurally unified} when one single trained system exposes multiple capabilities through shared interfaces, token spaces, or parameters. It is \emph{functionally unified} only when the resulting system provides an aggregate benefit without a practically meaningful loss in any explicitly declared essential capability.

This distinction is directly testable. Let $\mathcal{K}$ be a preregistered capability set, let $r_k(M)$ be its raw metric, and fix lower and upper anchors $(b_k,u_k)$ from the benchmark specification or a held-out calibration pool. After reversing lower-is-better metrics, define
\begin{equation}
s_k(M)=\operatorname{clip}_{[0,1]}
\left(\frac{r_k(M)-b_k}{u_k-b_k}\right).
\label{eq:normalize}
\end{equation}
Here $\operatorname{clip}_{[0,1]}$ maps values below or above the declared anchor range to 0 or 1. The anchors must be fixed before comparing systems, so normalization cannot be chosen to favor one model. Relative to a task-specialized or pre-unification reference $M_k^{\mathrm{ref}}$ evaluated under a matched inference budget, define
\begin{equation}
\begin{aligned}
\Delta_k&=s_k(M_{\mathrm{unified}})-s_k(M_k^{\mathrm{ref}}),\\
D&=\sum_{k\in\mathcal{K}} w_k\Delta_k ,
\end{aligned}
\label{eq:dividend}
\end{equation}
where $w_k\geq0$, $\sum_k w_k=1$, are reported rather than hidden. We call $D$ the \emph{unification dividend}. Given per-capability minimum practically important differences $\tau_k$ and confidence level $1-\alpha$, functional unification requires
\begin{equation}
\begin{aligned}
\operatorname{LCB}_{1-\alpha}(D)&\geq0,\\
\operatorname{LCB}_{1-\alpha}(\Delta_k)&\geq-\tau_k
\quad \forall k\in\mathcal{K},
\end{aligned}
\label{eq:functional}
\end{equation}
where confidence bounds are estimated by paired, task-stratified bootstrap resampling of reference and unified outputs, followed by recomputation of the normalized vector and dividend. The Unification Paradox is confirmed when $\operatorname{UCB}_{1-\alpha}(D)<0$ or $\operatorname{UCB}_{1-\alpha}(\Delta_k)<-\tau_k$ for any essential capability. A comparison that satisfies neither the functional-unification criterion nor this paradox criterion is \emph{inconclusive}. Capability sets, reference systems, weights, anchors, tolerances, budgets, and confidence levels must all be declared; they are deployment-specific rather than universal constants across settings.

The snapshot criterion can be complemented by an intervention test that directly exposes capability cannibalization. Let $M^{(0)}$ be a common checkpoint and $M^{(j)}$ the checkpoint after an intervention targeting capability $j$. Define the cross-capability effect matrix
\begin{equation}
C_{ij}=s_i(M^{(j)})-s_i(M^{(0)}).
\label{eq:effect-matrix}
\end{equation}
An intervention is capability-preserving when the lower confidence bound of its intended diagonal gain is positive and every off-diagonal lower bound satisfies its declared floor, $\operatorname{LCB}_{1-\alpha}(C_{ij})\geq-\tau_i$ for $i\neq j$. This matrix separates a merely co-located system from one in which improvements transfer or at least do not cannibalize other functions in practice.

As illustrated in \Cref{fig:overall}, next-generation UMMs should satisfy five fundamental full-stack \textit{unification guarantees}:

\vspace{-.1in}
\begin{enumerate}[leftmargin=2em, label=(\roman*).]
\setlength\itemsep{0em}

\item \textbf{Asymmetric Input Unification}:
UMMs must explicitly quantify modality utility and reliability across modalities rather than blindly aggregate heterogeneous evidence. (\S\ref{sec:asym-perception})

\item \textbf{Factorized Representation Unification}:
UMMs must allocate capacity across semantic abstraction and fine-grained reconstruction instead of forcing both into one latent space. (\S\ref{sec:representations})

\item \textbf{Stateful Reasoning Unification}:
UMMs must maintain evolving dynamic multimodal belief states beyond text-centric autoregressive outputs. (\S\ref{sec:reasoning})

\item \textbf{Synergistic Optimization Unification}:
Post-training must explicitly coordinate conflicting discriminative and generative objectives through robust gradient harmonization and Pareto-consistent reward shaping. (\S\ref{sec:posttraining})

\item \textbf{Multi-Centric Systemic Evaluation}:
Evaluation must measure systemic competence, capability preservation, and cross-task robustness rather than isolated leaderboard wins or single-benchmark optimality (\S\ref{sec:eval}).

\end{enumerate}

\vspace{-.1in}
The guarantees are coupled rather than a flat checklist: input calibration determines which evidence enters factorized representations; those representations support stable belief-state updates; optimization controls interference across the first three layers; and systemic evaluation closes the loop by detecting regressions and guiding redesign.

\noindent\textbf{Scope and staging.}
The evaluation guarantee should be implemented first because it supplies the preservation tests for every later intervention. Input calibration and representation factorization form the architectural foundation; persistent state is built on the evidence they preserve; and synergistic optimization is the transverse mechanism that coordinates all three. This dependency order narrows the roadmap into testable stages rather than five independent, equally immediate research programs.

\noindent\textbf{Relation to adjacent paradigms.}
Multimodal multitask learning and modality-competition work primarily study parameter sharing and token-level dominance; gradient surgery and multi-objective optimization address conflicts among specified losses~\citep{yu2020gradient,sener2018multi}; and holistic benchmarks aggregate broad but often task-isolated scores. We do not claim that these individual ingredients are new. Our contribution is to connect them through one cross-layer diagnosis, an explicit capability-preservation criterion, and a dependency-aware roadmap for functional unification.
\vspace{-15pt}

\section{Asymmetric Input Unification}
\label{sec:asym-perception}
\vspace{-10pt}

\begin{wrapfigure}{r}{0.5\textwidth}
    \centering
    \vspace{-.1in}
    \includegraphics[width=\linewidth]{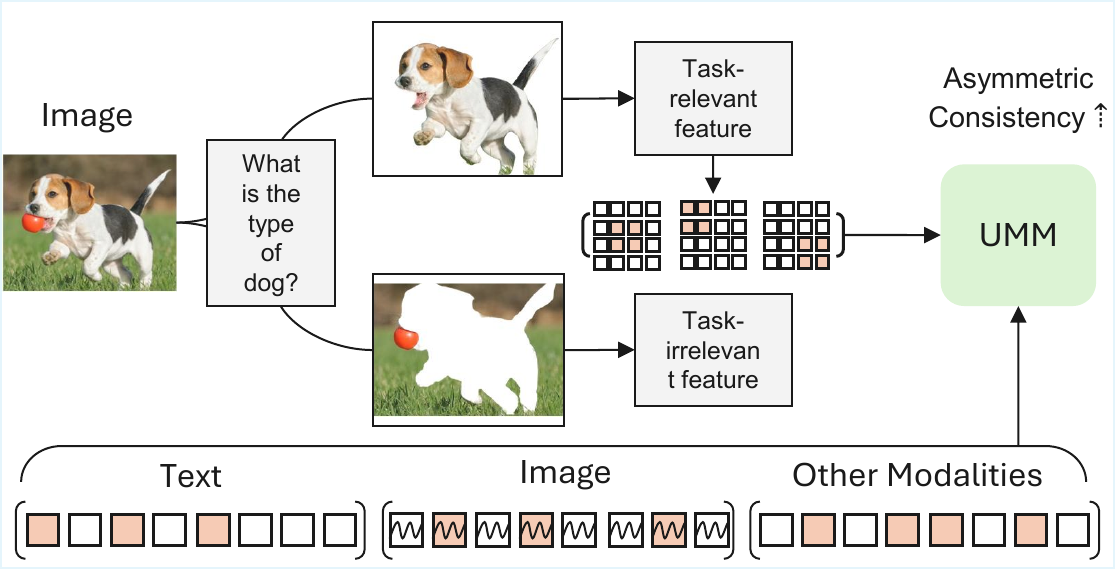}
    \caption{The input-level unification paradox: structurally homogeneous processing ignores asymmetric evidence quality, allowing unreliable modalities to corrupt unified internal states.}
    \label{fig:perception}
    \vspace{-.15in}
\end{wrapfigure}

The first fracture of structural unification appears at the input level.
Current UMMs often treat input unification as formatting: concatenate features, tokenize every modality into a shared sequence, or pass them through a fixed interface.
Yet multimodal inputs exhibit inherent \textbf{evidence asymmetry} (\Cref{fig:perception}): modalities differ in information density, reliability, completeness, and task relevance~\citep{he2024efficient, zhang2024multimodal}.
Thus, input unification should be dynamic evidence calibration rather than blind feature aggregation.
\vspace{-5pt}

\subsection{Challenges}
\label{sec-input-challenge}
\vspace{-5pt}

\textbf{The input-level unification paradox.}
Naive structural unification enforces homogeneous processing over heterogeneous evidence.
Equal token allocation, uniform cross-attention, or fixed fusion pathways implicitly treat all modalities as similarly informative, allowing unreliable evidence to contaminate shared states.
This yields two failures: adding modalities may \emph{degrade} performance when noisy or irrelevant inputs inject alignment noise~\citep{huang2022modality}, and incomplete or contradictory evidence may trigger overconfident hallucination rather than calibrated uncertainty~\citep{li2023pope, zollo2025confidence, zitkovich2023rt}.
Both arise because current UMMs unify interfaces without estimating evidence utility.
\vspace{-5pt}

\subsection{Research Directions}
\label{sec-input-direction}
\vspace{-5pt}

\textbf{Dynamic input unification.}
True input unification requires task-aware utility modeling.
A unified model should estimate which evidence is reliable, missing, or useful for the current objective; route unreliable signals away from shared states; and abstain when evidence is insufficient or mutually inconsistent.
Because utility is contextual rather than intrinsic, this calibration must depend on the task, modality, and evidence state.
The desired behavior is evidence-sensitive coordination: performance should improve when informative evidence is added, remain stable under irrelevant evidence, and degrade gracefully when critical evidence is missing.
Missing evidence should not be treated as an ordinary token pattern to be imputed away: the model should propagate an explicit reliability state into later representation, reasoning, and generation modules.
If the model is already confused about what evidence it has, later layers cannot reliably recover functional coherence.

\noindent\textbf{A concrete utility-gated instantiation.}
A reliability head scores each modality from its encoded evidence and observable quality metadata, while a query-conditioned utility head estimates relevance to the current task. Their scores determine fusion weights and are retained as an explicit uncertainty state for later modules. Counterfactual modality deletion and controlled corruption provide supervision: predicted utility should track the change in task loss when evidence is removed, and predicted reliability should track known corruption. At inference, unreliable modalities are routed to capability-specific adapters or excluded, while contradictory high-utility evidence triggers abstention. This proposal is tested through utility-calibration error, clean-to-corrupt performance slope, and false-confidence rate under realistic missing or conflicting modalities.

\noindent
\textbf{Faithful generation under asymmetric evidence.}
Generation should condition on calibrated evidence states rather than fused vectors of unknown reliability.
When visual input is noisy, text is underspecified, or modalities conflict, the model should avoid inventing unsupported details; calibrated abstention is therefore also a generative safety mechanism.

Existing ingredients point in this direction: token pruning and merging estimate patch utility~\citep{rao2021dynamicvit, ryoo2021tokenlearner, bolya2023tome}, mixture-of-experts enables conditional routing~\citep{fedus2022switch}, and selective prediction motivates risk-aware abstention~\citep{geifman2019selectivenet}.
Faithfulness benchmarks such as POPE and HallusionBench~\citep{li2023pope, guan2024hallusionbench}, together with studies of language priors~\citep{agrawal2018gvqa, ramakrishnan2018adversarial_vqa}, show why such utility estimates must target semantic reliability rather than efficiency alone.
\vspace{-5pt}

\section{Factorized Representation Unification}
\label{sec:representations}
\vspace{-5pt}

If input-level unification determines what evidence enters the model, representation-level unification determines what information can be faithfully preserved, transformed, and shared across capabilities.
This is the core paradox of UMMs.
Current models often assume that one shared latent space can simultaneously support understanding, reasoning, and generation, yet these capabilities impose conflicting representational requirements~\citep{chen2025janus,team2024chameleon,luo2026flexumm,luo2026latentumm}.
The question is therefore what kind of representation can unify heterogeneous objectives without destructive interference.

\vspace{-5pt}
\subsection{Challenges}
\label{sec-repre-challenge}
\vspace{-5pt}

\textbf{The representation unification paradox.}
Forcing all modalities and tasks into one latent space creates a clean interface but collapses two incompatible pressures.
Understanding favors lossy semantic abstraction, while generation requires fine-grained reconstruction of layout, texture, color, and boundaries~\citep{chow2024unified}.
If the representation is too semantic, details vanish; if it preserves too much low-level information, semantic reasoning becomes noisy.
Scale cannot recover information erased at the encoding stage exposed in recent systems: decoupled pathways reduce capacity conflict but risk cross-branch inconsistency~\citep{chen2025janus, chen2024janus, deng2025emerging, fan2025unified}, whereas highly unified representations are efficient but often imbalanced~\citep{team2024chameleon, xie2024show}.
Chameleon discards fine visual detail during discretization~\citep{team2024chameleon}, and Tuna improves alignment while still leaning toward semantic encoders~\citep{liu2025tuna}.
Thus, ideal representation unification is capacity allocation, not simple vector mixing.

\vspace{-5pt}
\subsection{Research Directions}
\label{sec-repre-direction}
\vspace{-5pt}
\textbf{From shared latent spaces to factorized unification.}
True representation unification should coordinate two roles rather than force them into one latent: a semantic stream retains task-sufficient invariants for abstraction and grounding, while a reconstructive stream preserves spatial and perceptual detail. They are trained with their corresponding semantic and reconstruction objectives, plus a bottleneck or routing constraint that prevents either stream from becoming an unrestricted duplicate~\citep{tishby2000information}. A selective bridge should transmit identity, relations, and intent from the semantic stream and expose local geometry from the reconstructive stream only when requested.

\noindent\textbf{One architectural realization.}
A dual-stream encoder writes global and object-level tokens to the semantic stream and higher-resolution residual patches to the reconstructive stream. A query-conditioned gate controls how much residual detail enters a given understanding, editing, or grounded-generation head. Alternating semantic, reconstruction, and bridge batches limits shortcut learning; held-out linear probes and round-trip reconstruction determine whether each stream retains its assigned information. This is an implementable candidate, not a uniquely correct architecture.

\noindent\textbf{Measurable capacity allocation.}
A factorized representation should retain semantic invariance under texture or style changes and reconstructive fidelity for layout, boundaries, and fine attributes~\citep{cover2006elements,yu2024representation}. Controlled-interference probes should test both directions: generation-oriented updates should not degrade semantic probes, and semantic updates should not erase reconstructive detail. Decoupled architectures and recent alignment methods are partial instances of this design space~\citep{chen2024janus,chen2025janus,deng2025emerging,liu2025tuna,xie2025reconstruction}; our proposal is the preservation criterion and bridge design, not a claim that factorization itself is new.

\vspace{-.15in}
\section{Stateful Reasoning Unification}
\label{sec:reasoning}

While unified multimodal models (UMMs) have achieved promising capabilities in multimodal understanding and generation, their reasoning process remains fundamentally \textit{text-centric}.
Current UMMs reason through autoregressive token generation, where non-text modalities serve as conditional signals rather than active participants.
Reasoning intelligence is therefore reflected in final text, while internal trajectories remain implicit and transient.

\begin{wrapfigure}{r}{0.5\textwidth}
    \vspace{-0.3in}
    \centering
    \includegraphics[width=\linewidth]{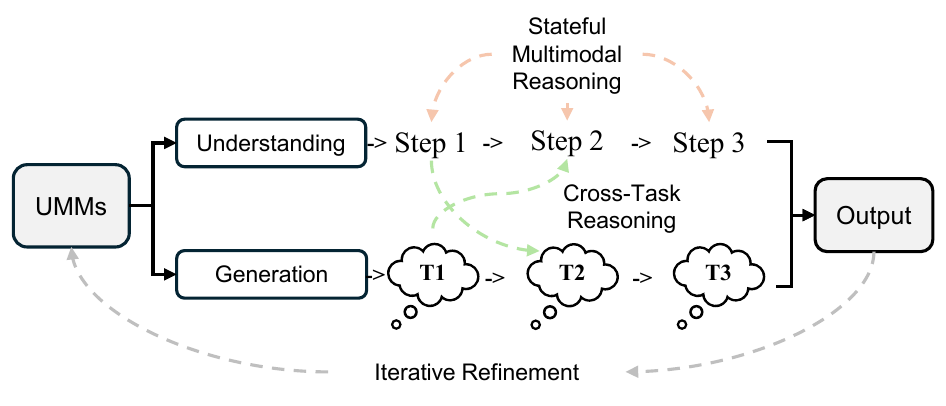}
    \caption{
    The reasoning unification paradox in UMMs.
    Although current UMMs operate under unified multimodal architectures, reasoning remains text-centric and autoregressive, lacking a persistent internal belief state across tasks.
    We envision stateful reasoning unification, where multimodal reasoning evolves through explicit, revisable, and cross-task consistent belief structures.
    }
    \label{fig:reasoning}
    \vspace{-0.2in}
\end{wrapfigure}

This exposes a fundamental \textbf{reasoning unification paradox}: although UMMs are trained under a unified architecture and latent space, their reasoning process is still fragmented into task-local autoregressive flows.
Intermediate states are not explicitly preserved, revised, or shared across interactions: explanation, editing, and dialogue over the same scene are often treated as independent calls.
Consistency is enforced at the output level rather than through persistent internal cognition.

True multimodal reasoning therefore requires \textbf{stateful reasoning unification}: persistent belief structures that can be accumulated, updated, verified, and geometrically anchored across modalities and tasks (\Cref{fig:reasoning}).

\vspace{-.15in}
\subsection{Challenges}

The challenge is the mismatch between shared multimodal representations and unidirectional reasoning.
Intermediate inferences are encoded only implicitly in generated tokens, so they are hard to explicitly preserve, revise, or reconcile with later evidence.
In multi-step settings spanning understanding, explanation, editing, retrieval, and dialogue, this produces accumulated hallucinated revisions and cross-modal task-level contradictions.

\vspace{-.15in}
\subsection{Research Directions}

\textbf{Persistent Multimodal Belief States.}
All physical carriers should implement the same state interface. Given calibrated evidence $e_t$, the previous state $b_{t-1}$, and the last action $a_{t-1}$, the model proposes $\tilde b_t=U_\theta(b_{t-1},e_t,a_{t-1})$. A gate that accounts for uncertainty and contradiction then accepts or rejects the proposal:
\begin{equation}
c_t=\sigma C_\theta(b_{t-1},e_t),\quad b_t=c_t\odot\tilde b_t+(1-c_t)\odot b_{t-1},\quad \hat y_t=H_\theta(b_t,e_t).
\label{eq:belief}
\end{equation}
Here $U_\theta$ proposes a new belief, $c_t$ is a dimension-wise acceptance gate, and $H_\theta$ reads the accepted state to produce the next output. When the evidence is reliable and consistent, $c_t$ admits the proposal; when it is unreliable or contradictory, the old state is preserved until revision is justified. Training combines the task objective with three auxiliary signals: cycle consistency asks the state to support re-understanding after generation or editing; contradiction supervision teaches the gate to preserve, revise, or roll back commitments as evidence changes; and a sparse-revision penalty discourages unnecessary state churn.

\begin{table}[t]
\vspace{-5pt}
\centering
\footnotesize
\setlength{\tabcolsep}{3pt}
\begin{tabular}{@{}p{0.20\linewidth}p{0.38\linewidth}p{0.34\linewidth}@{}}
\toprule
\textbf{Carrier} & \textbf{Physical state and update} & \textbf{Best fit and main risk} \\
\midrule
Latent state-space track & Fixed recurrent latent slots updated beside token decoding by Eq.~\ref{eq:belief}. & Efficient dense dynamics; difficult to inspect and geometrically ground. \\
Memory registers & Persistent key--value tokens with entity, provenance, time, and confidence fields; gated writes and retrieval. & Drop-in multi-turn memory; fixed capacity can overwrite old commitments. \\
Dynamic scene graph & Entity nodes and typed spatial/causal edges updated by detection and message passing. & Explicit geometry and rollback; brittle entity matching can fragment state. \\
\bottomrule
\end{tabular}
\caption{Three implementable carriers for a persistent multimodal belief state. All share the update interface and supervision signals but make different structural commitments.}
\label{tab:belief-carriers}
\vspace{-.2in}
\vspace{-5pt}
\end{table}

\vspace{-5pt}
These are architectural candidates, not validated prescriptions. In each case, commitments formed during understanding constrain later explanation, editing, and generation, while provenance, uncertainty, and selective rollback permit revision rather than irreversible forward generation. A decisive experiment reuses the same asset across an understand--explain--edit--re-understand sequence and measures state contradiction, identity retention, rollback accuracy, and task performance against an otherwise matched stateless decoder. Existing multimodal reasoning remains largely autoregressive and task-local~\citep{zhang2025unified2,wang2025multimodal,yan2025survey,wang2024measuring,li2025visiomath}; multimodal self-correction is still limited when state must persist across tasks~\citep{madaan2023self,ji2025reflair,luo2025coherent}.

\vspace{-5pt}
\section{Synergistic Optimization Unification}
\label{sec:posttraining}

UMM post-training combines discriminative, autoregressive, diffusion, and preference objectives~\citep{christiano2017deep,ouyang2022training,rafailov2023direct}. Their gradients and reward proxies can conflict, so an update that improves one capability may damage another. \textbf{Synergistic optimization unification} therefore treats the capability interaction, rather than a single task loss, as the unit of alignment (Appendix Figure~\ref{fig:posttrain}).

\vspace{-.05in}
\subsection{Challenges}
\label{sec-post-challenge}
\vspace{-5pt}

\textbf{Gradient and reward conflict.}
Understanding emphasizes semantic accuracy, while generation emphasizes realism, diversity, and reconstruction. Isolated rewards score local quality rather than compatibility across understanding, generation, editing, and reasoning; uniform sharing can therefore cause negative transfer, whereas complete isolation loses useful transfer.
\vspace{-5pt}

\subsection{Research Directions}
\label{sec-post-direction}

\textbf{Conflict-aware gradient routing.}
At each update, the model computes a gradient for every capability. Components with non-negative cosine agreement update shared parameters; conflicting components are projected away from one another or routed to capability-specific adapters, following established gradient-surgery or multi-objective methods~\citep{yu2020gradient,sener2018multi}. A small Pareto solver can upweight capabilities near their preservation floor. This specifies the coordination logic without presenting one unvalidated projection formula as a new optimizer.

\noindent\textbf{Optimization procedure.}
Each tentative update applies the chosen projection or routing rule to shared parameters, routes conflicting residuals to capability-specific adapters, and is checked on held-out capability sentinels. It is accepted only when the target improves and every lower confidence bound remains above its preservation floor; otherwise the step, weights, or routing are revised. Appendix~\ref{app:optimization-procedure} gives the full six-step procedure.

\noindent\textbf{Pareto-consistent rewards.}
Reward models should score local quality together with cross-capability preservation: an image should be plausible and respect inferred spatial relations; an answer should be correct and preserve evidence needed for later editing. Existing UMM alignment efforts begin to combine such objectives~\citep{su2025unigame,xie2025reconstruction,yan2025unified,jiang2025t2i,bai2026unipath}, while the open problem is to coordinate their gradients and measure collateral capability movement.
\vspace{-5pt}

\section{Multi-Centric Systemic Evaluation}
\label{sec:eval}
\vspace{-5pt}

Evaluation is where the Unification Paradox becomes observable. Isolated task scores can make structural unification appear successful while collateral capability losses remain hidden. Evaluation must therefore measure whether understanding, generation, reasoning, grounding, editing, and safety remain mutually compatible after an architectural or training intervention.

\subsection{Challenges}
\label{sec-eval-challenge}

\textbf{Fragmented metrics and evidence.}
VQA accuracy, language-generation scores~\cite{papineni2002bleu}, image-generation metrics~\cite{heusel2017gans}, and reasoning diagnostics measure local skill rather than mutual compatibility. Existing studies nevertheless provide convergent case evidence: structurally unified and decoupled designs expose different semantic--reconstructive trade-offs~\citep{team2024chameleon,chen2024janus,chen2025janus}; alignment interventions can shift the balance between understanding and reconstruction~\citep{liu2025tuna,xie2025reconstruction}; and TorchUMM reports non-monotonic scaling and cross-capability movement~\citep{luo2026torchumm}. Because these works use heterogeneous models, data, and metrics, we treat them as a structured evidence map rather than pool them into a statistically comparable meta-analysis.

\noindent\textbf{Quantitative failure signatures.}
The paired post-training movements reported in the Introduction are one direct signature of capability cannibalization. Joint benchmarks reveal a second. MME-Unify evaluates 12 models and reports that, on commonsense question answering with both reasoning and image requirements, the strongest open-source model obtains $59.65\%$ accuracy but only $38\%$ accuracy-plus; no open-source model exceeds $30\%$ on its other unified tasks, and none completes the multi-step Visual-CoT task~\citep{xie2025mme}. It also reports a model reaching $76.24\%$ text-reasoning accuracy while failing to produce a correct paired image. Uni-MMMU exposes bidirectional, process-level dependencies across eight reasoning-centric domains~\citep{zou2025uni}. These are observable output disagreement, cascading state error, and collateral regression, not controlled estimates of structural sharing. Appendix Table~\ref{tab:benchmark-gaps} systematically contrasts the benchmark scopes and the additional interventional question posed here.

\subsection{Research Directions}
\label{sec-eval-direction}
\vspace{-5pt}

\noindent\textbf{Capability matrix.}
Every intervention is audited across understanding, generation, reasoning/state, editing, and safety/hallucination rather than only its target task. Appendix Table~\ref{tab:preservation-details} specifies the metric families and cross-task preservation probes for all five rows.

\noindent\textbf{Reading the intervention matrix.}
The diagonal of Eq.~\ref{eq:effect-matrix} records an intervention's intended gain, while its off-diagonal entries record side effects. A negative row identifies a vulnerable capability; a negative column identifies a cannibalizing intervention; and positive off-diagonal structure reveals transfer. These patterns guide whether to recalibrate inputs, separate representational capacity, revise state updates, or route conflicting gradients.

\noindent\textbf{Protocol and reporting.}
First, preregister the capability set, matched reference models or checkpoints, test items, raw metrics and anchors, inference budget, weights, tolerances $\{\tau_k\}$, and confidence level. Second, evaluate the common checkpoint and each intervention on the complete matrix using the same items and decoding budget. Third, orient and normalize scores with Eq.~\ref{eq:normalize}; compute $\Delta_k$, $D$, and the appropriate column of $C$; and obtain paired bootstrap confidence intervals. Finally, report the full $\mathbf{s}$ and $\boldsymbol{\Delta}$ vectors, their worst component, mean, variance, and the dividend $D$. These summaries support comparison, while Eq.~\ref{eq:functional} prevents a favorable mean from hiding one collapsed capability. A result is labeled functionally unified, paradoxical, or inconclusive according to those confidence bounds. Pareto curves and negative-transfer coefficients can supplement, not replace, the vector and effect-matrix reports. This protocol is actionable without assuming that its scalar summaries are universal fixed constants.

\noindent\textbf{Cross-task and longitudinal probes.}
Latent probes test semantic and reconstructive retention, while round-trip generation and editing expose state drift~\citep{aravindan2025vlms,goel2022cyclip,zhang2024cross,kong2025round,xie2025reconstruction}. Mixed-modality benchmarks provide diverse evaluation tasks~\citep{zou2025uni,xie2025mme,liu2025umnibench,wei2025ggbench,li2025gir,wu2025omnigen2,niu2025does}; the paired protocol asks what changed elsewhere after one capability improved.

\bibliographystyle{plainnat}
\bibliography{references}

\clearpage
\appendix

\section{Supplementary Roadmap Details}
\label{app:benchmark-comparison}

\setcounter{table}{0}
\renewcommand{\thetable}{A\arabic{table}}
\begin{table}[t]
\centering
\footnotesize
\setlength{\tabcolsep}{4pt}
\begin{tabular}{@{}p{0.15\linewidth}p{0.28\linewidth}p{0.49\linewidth}@{}}
\toprule
\textbf{Benchmark} & \textbf{What it establishes} & \textbf{Gap relative to the Unification Paradox} \\
\midrule
MME~\citep{xie2025mme}, MMBench~\citep{liu2024mmbench}, SEED-Bench~\citep{ge2024seed} & MME and MMBench emphasize perception and understanding; SEED-Bench expands hierarchical modality and task coverage. & Their absolute task scores do not require a paired pre/post trace, matched references, or preservation floors; they therefore cannot attribute collateral movement to a unification intervention. \\
MME-Unify~\citep{xie2025mme} & Standardized understanding and generation plus five mixed-modality tasks over one evaluation suite. & Stronger joint coverage, but the aggregate is a model snapshot rather than a paired pre/post capability vector; it imposes no matched reference or mandatory per-capability floor, and its authors note that simplified multiple-choice/CLIP scoring can inflate some results. \\
Uni-MMMU~\citep{zou2025uni} & Eight bidirectionally coupled reasoning tasks with intermediate/final and text/image evaluation. & Diagnoses synergy within its reasoning-centric domains, but does not require an intervention trace or preservation checks over unrelated base capabilities, reliability perturbations, hallucination, and safety. \\
Our proposed matrix & Matched baseline/intervention evaluation over understanding, generation, reasoning, editing, and safety. & Adds normalized effect vector $C_{\cdot j}$, uncertainty, aggregate dividend, and mandatory preservation floors; it complements rather than replaces task benchmarks. \\
\bottomrule
\end{tabular}
\caption{Existing benchmarks answer progressively stronger questions, but an interventional preservation protocol is needed to diagnose capability cannibalization. ``Gap'' denotes a difference in evaluation target, not a claim that a benchmark fails at its stated purpose.}
\label{tab:benchmark-gaps}
\end{table}

\begin{table}[t]
\centering
\footnotesize
\setlength{\tabcolsep}{4pt}
\begin{tabular}{@{}p{0.16\linewidth}p{0.26\linewidth}p{0.50\linewidth}@{}}
\toprule
\textbf{Capability} & \textbf{Metric family} & \textbf{Preservation and consistency probe} \\
\midrule
Understanding \& grounding & Accuracy, calibration, grounding quality & Add, remove, or corrupt evidence; informative evidence should help, while conflicts should increase uncertainty rather than hallucination. \\
Generation \& reconstruction & Perceptual quality, constraint satisfaction & Compare fidelity before and after understanding-oriented updates; verify identity, layout, and stated relations. \\
Reasoning/state & Task accuracy, contradiction rate & Reuse the same asset across explanation and dialogue; test belief updates, rollback, and multi-turn consistency. \\
Editing & Instruction and identity preservation & Run understand--edit--re-understand loops and measure whether requested changes occur without collateral drift. \\
Safety/hallucination & Hallucination and abstention rates & Perturb modality reliability and test whether confidence tracks available evidence~\citep{li2023pope,guan2024hallusionbench}. \\
\bottomrule
\end{tabular}
\caption{Detailed capability-preservation matrix. Each intervention is evaluated across the same declared capability set, not only on its target task.}
\label{tab:preservation-details}
\end{table}

\subsection{Illustrative Optimization Procedure}
\label{app:optimization-procedure}

At each update: (1) sample one mini-batch per capability and compute $g_k$; (2) estimate the cosine-conflict matrix; (3) apply the chosen projection or routing rule to shared parameters and send remaining residuals to the corresponding adapters; (4) tentatively update the model; (5) evaluate lightweight held-out sentinel sets and populate one column of $C$ in Eq.~\ref{eq:effect-matrix}; and (6) accept the update only if the target improves and all lower confidence bounds remain above their preservation floors; otherwise reduce the step, change the capability weights, or increase routing. Full matrix evaluation is periodic rather than per-step. This procedure makes the acceptance logic verifiable while leaving the choice of gradient surgery or Pareto solver open~\citep{yu2020gradient,sener2018multi}.

\begin{figure}[b]
    \centering
    \includegraphics[width=.8\linewidth]{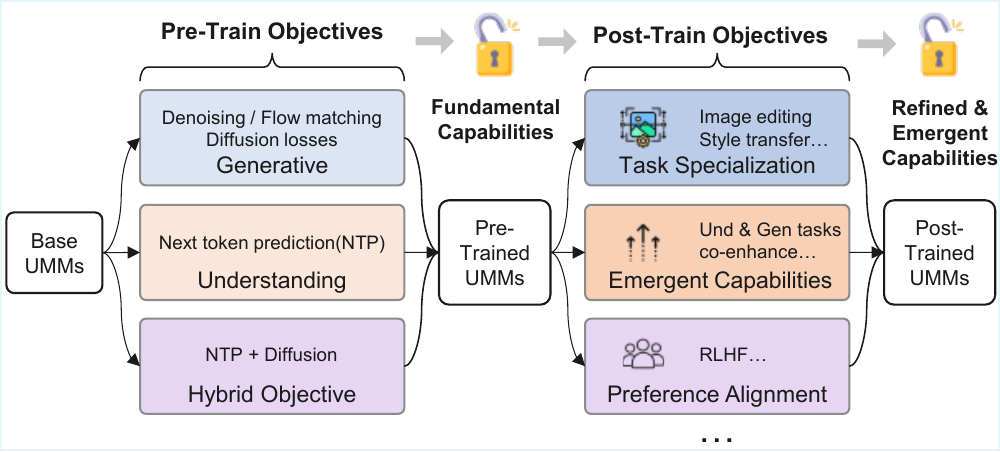}
    \caption{The optimization unification paradox. Heterogeneous post-training objectives can cause gradient conflict and negative transfer; coordinated gradients and rewards instead target robust joint capability.}
    \label{fig:posttrain}
\end{figure}

\section*{Limitations}

This paper is a position and roadmap contribution, not an empirical benchmark or a validated end-to-end UMM architecture. The proposed input gate, factorized bridge, belief-state carriers, and gradient-routing procedure are illustrative instantiations whose effectiveness, stability, and computational cost must be tested in future systems. The quantitative values reported here are a secondary synthesis of published evaluations, not new experiments. Those studies include heterogeneous architectures, datasets, and metrics; they expose failure signatures but do not establish a universal causal relationship between structural sharing and capability interference. A statistically pooled meta-analysis would be misleading until models are evaluated under a common protocol, and even paired post-training results attribute effects to the complete intervention rather than to one component in isolation.

The operational definitions also require deployment-specific choices. Capability sets, reference systems, inference budgets, normalization anchors, weights, preservation tolerances, and confidence levels can change the unification dividend and preservation decision. Reporting the full capability and intervention vectors, confidence intervals, and sensitivity analyses is therefore essential. The proposed matrix also inherits coverage gaps from its constituent tasks and cannot certify behavior outside the declared capability set. Finally, building standardized cross-task assets and longitudinal probes requires community engineering beyond this conceptual blueprint. The primary contribution of this paper is the diagnosis, falsifiable criteria, and roadmap; it does not claim that functional unification has already been achieved.

\section{Call to Action}
UMMs should be judged by functional, not merely structural, unification. The central lesson is not that sharing is undesirable, but that sharing is a hypothesis whose system-wide effects must be measured. Structural consolidation is useful when it enables transfer without erasing information, destabilizing persistent state, or hiding losses in another capability. When interference persists, selective specialization and explicit coordination are more credible than uniform sharing by default.

The immediate priority is a common audit protocol. Researchers should preregister the capability set, references, budgets, and preservation thresholds; evaluate matched checkpoints on shared items; and publish the full capability vectors and intervention matrix with uncertainty. Only after this baseline is available can input calibration, representation factorization, stateful reasoning, and gradient routing be compared meaningfully. We therefore propose a staged agenda: establish evaluation first; test evidence calibration and capacity allocation; evaluate persistent state on longitudinal tasks; and then coordinate gradients and rewards across the validated components.

Each mechanism should be introduced as a bounded intervention with a matched alternative and a declared failure condition. Full sharing should remain the simpler choice when the intervention matrix shows neutral or positive transfer; additional routing, memory, or factorization is justified only by measured interference. Progress therefore requires more than isolated leaderboard gains: it requires a non-negative aggregate dividend together with satisfied capability-preservation floors. A genuinely unified model should improve capabilities jointly---or expose uncertainty and abstain---while keeping its cross-capability effects visible and reproducible.

\end{document}